\documentclass{article}
\usepackage{isit2004}  

\usepackage{amsmath}

\begin{document}
\topmargin = 0mm

\itwtitle{The Private Key Capacity Region for Three Terminals}

 \itwauthor{Chunxuan Ye\ \ \ and\ \ \ Prakash Narayan}
           {Department of Electrical and Computer Engineering\\
            University of Maryland \\
            College Park, MD 20742, U.S.A.\\
            e-mail: {\tt \{cxye,\ prakash\}@eng.umd.edu}
           }

\itwmaketitle

\begin{itwabstract}
We consider a model with three terminals and examine the problem of characterizing 
the largest rates at which two pairs of terminals can simultaneously generate 
private keys, each of which is effectively concealed from the remaining terminal. 

\end{itwabstract}

\begin{itwpaper}

\itwsection{Introduction}  

Suppose that terminals ${\cal X}$, ${\cal Y}$ and ${\cal Z}$ observe, respectively, 
the distinct components of a discrete memoryless multiple source, i.e., i.i.d. 
repetitions of the random variables (rvs) $X$, $Y$, $Z$, respectively. The terminals 
are permitted unrestricted communication among themselves over a public channel, and 
all the transmissions are observed by all the terminals. An eavesdropper has access 
to this public communication but gathers no additional side information; also, the 
eavesdropper is passive, i.e., unable to corrupt the transmissions. Terminals 
${\cal X}$ and ${\cal Y}$ (resp. ${\cal X}$ and ${\cal Z}$) generate a ``private key'' 
(PK) with the possible help of terminal ${\cal Z}$ (resp. ${\cal Y}$) which is 
concealed from the helper terminal ${\cal Z}$ (resp. ${\cal Y}$) and from an 
eavesdropper with access to the public communication among the terminals. Our main 
technical results are single-letter outer and inner bounds for the PK-capacity 
region. Further, under certain special conditions, for instance if the correlation 
of $Y$ and $Z$ is deterministic (i.e., there exists a common function of $Y$ and $Z$ 
which renders them conditionally independent), these bounds coincide to yield the 
PK-capacity region.

\itwsection{Statement of Results}

Consider a discrete memoryless multiple source (DMMS) with three components 
corresponding to generic rvs $X$, $Y$, $Z$ with finite alphabets ${\cal X}$, 
${\cal Y}$, ${\cal Z}$. Let $X^n=(X_1,\cdots ,X_n)$, $Y^n=(Y_1,\cdots , Y_n)$, 
$Z^n=(Z_1,\cdots , Z_n)$ be $n$ i.i.d. repetitions of the rvs $X$, $Y$, $Z$. 
The terminals ${\cal X}$, ${\cal Y}$, ${\cal Z}$ respectively observe the 
components $X^n$, $Y^n$, $Z^n$ of the DMMS $(X^n, Y^n, Z^n)$, where $n$ 
denotes the observation length. The terminals can communicate with each other 
through broadcasts over a noiseless public channel, possibly interactively in 
many rounds. Following \cite{CsiNar03}, we assume without loss of generality 
that these transmissions occur in consecutive time slots in $r$ rounds; the 
communication is depicted by $3r$ rvs $F_1, \cdots ,F_{3r}$, where $F_t$ 
denotes the transmission in time slot $t$, $1\leq t\leq 3r$, by a terminal 
assigned an index $i=t\mod 3$, $1\leq i\leq 3$, with terminals ${\cal X}$, 
${\cal Y}$, ${\cal Z}$ corresponding to indices 1, 2, 3, respectively. In 
general, $F_t$ is allowed to be any function, defined in terms of a mapping 
$f_t$, of the observations at the terminal with index $i$ and of the previous 
transmissions $F^{t-1}=(F_1,\cdots, F_{t-1})$. Let ${\bf F}$ denote 
collectively all the transmissions. Randomization at the terminals is not 
permitted. 

The rvs $K_{{\cal XY}}$, $L_{{\cal XY}}$ represent an $\varepsilon$-private 
key ($\varepsilon$-PK) for the terminals ${\cal X}$ and ${\cal Y}$ which is 
private from the helper terminal ${\cal Z}$, achievable with communication 
${\bf F}$, if $K_{{\cal XY}}$ and $L_{{\cal XY}}$ are functions of the data 
available at terminals ${\cal X}$ and ${\cal Y}$, respectively, i.e., 
$K_{{\cal XY}}=K_{{\cal XY}}(X^n, {\bf F}),\ L_{{\cal XY}}=L_{{\cal XY}}(Y^n, 
{\bf F})$; $K_{{\cal XY}}$ and $L_{{\cal XY}}$ take values in the same finite 
set ${\cal K_{XY}}$ with $\Pr\{K_{{\cal XY}}\neq L_{{\cal XY}}\}\leq \varepsilon$; 
$K_{{\cal XY}}$ (or $L_{{\cal XY}}$) satisfies the {\it secrecy condition} 
$\frac{1}{n}I(K_{{\cal XY}}\wedge {\bf F}, Z^n)\leq \varepsilon$; and 
$K_{{\cal XY}}$ (or $L_{{\cal XY}}$) satisfies the {\it uniformity condition} 
$\frac{1}{n}H(K_{{\cal XY}})\geq \frac{1}{n}\log |{\cal K_{XY}}|- \varepsilon$. 
We are interested in the {\it simultaneous generation} of individual PK pairs 
($K_{{\cal XY}}$, $K_{{\cal XZ}}$) as above.

A pair of numbers ($R_{{\cal XY}}$, $R_{{\cal XZ}}$) is an {\it achievable 
PK-rate pair} if $\varepsilon_n$-PK pairs ($K_{{\cal XY}}^{(n)}$, 
$K_{{\cal XZ}}^{(n)}$) are achievable with suitable communication, such that 
$\varepsilon_n\rightarrow 0$, 
$\frac{1}{n}H(K_{{\cal XY}}^{(n)})\rightarrow R_{{\cal XY}}$, 
$\frac{1}{n}H(K_{{\cal XZ}}^{(n)})\rightarrow R_{{\cal XZ}}$. The set of all 
achievable PK-rate pairs is the {\it PK-capacity region} $C_{PK}$.

Our main results for the PK-capacity region are the following.

{\bf Theorem 1} ({\it Outer bound for $C_{PK}$}): Let ($R_{\cal XY}$, 
$R_{\cal XZ}$) be an achievable PK-rate pair. Then
\begin{equation}
R_{{\cal XY}}\leq I(X\wedge Y|Z), \ \ \ \ \  R_{\cal XZ}\leq I(X\wedge Z|Y),
\label{e5}
\end{equation}
\[
R_{\cal XY}+R_{\cal XZ}\leq I(X\wedge Y,Z)-
\max_{U:\ U -\!\!\circ\!\!- Y -\!\!\circ\!\!- XZ,\ U -\!\!\circ\!\!- Z 
-\!\!\circ\!\!- XY}I(U\wedge X).
\]

{\bf Theorem 2} ({\it Inner bound for $C_{PK}$}): The PK-capacity region 
$C_{PK}$ is inner-bounded by the convex hull of the regions
\[
  \left\{\begin{array}{ll}
    (R_{\cal XY}, R_{\cal XZ}): & 0\leq R_{\cal XY}\leq 
I(X\wedge Y|U_{mss(Y)},Z),\\
                                & 0\leq R_{\cal XZ}\leq 
I(X\wedge Z|Y), \\
                                & R_{\cal XY}+R_{\cal XZ}\leq 
I(X\wedge Y,Z)-I(U_{mss(Y)}\wedge X)
                 \end{array}
          \right\}
\]
and
\[
  \left\{\begin{array}{ll}
    (R_{\cal XY}, R_{\cal XZ}): & 0\leq R_{\cal XY}\leq I(X\wedge Y|Z),\\
                                & 0\leq R_{\cal XZ}\leq 
I(X\wedge Z|V_{mss(Z)}, Y), \\
                                & R_{\cal XY}+R_{\cal XZ}\leq 
I(X\wedge Y,Z)-I(V_{mss(Z)}\wedge X)
                 \end{array}
          \right\},
\]
where $U_{mss(Y)}$ (resp. $V_{mss(Z)}$) is the {\it minimal sufficient 
statistic} for $Y$ (resp. $Z$) w.r.t. $Z$ (resp. $Y$).

{\bf Theorem 3:} If there exists a rv $U$ such that 
\begin{equation}
U -\!\!\circ\!\!- Y -\!\!\circ\!\!- XZ, \ \ \ \ \ U -\!\!\circ\!\!- Z 
-\!\!\circ\!\!- XY,\ \ \ \ \ Y -\!\!\circ\!\!- U -\!\!\circ\!\!- Z,
\label{e13}
\end{equation}
the PK-capacity region equals the set of pairs ($R_{\cal XY}$, 
$R_{\cal XZ}$) which satisfy (\ref{e5}) and 
\[
R_{\cal XY}+R_{\cal XZ}\leq I(X\wedge Y,Z)-\max_{U} I(U\wedge X),
\]
where the maximum is w.r.t. $U$ satisfying (\ref{e13}).

{\bf Theorem 4:} If $Y$ and $Z$ are {\it deterministically correlated}, 
the PK-capacity region $C_{PK}$ equals the set of pairs ($R_{\cal XY}$, 
$R_{\cal XZ}$) which satisfy (\ref{e5}) and 
\[
R_{\cal XY}+R_{\cal XZ}\leq I(X\wedge Y,Z)-I(U_{mcf}\wedge X),
\]
where $U_{mcf}$ is the {\it maximal common function} of $Y$ and $Z$.

\end{itwpaper}

\begin{itwreferences}

\bibitem{CsiNar03}
I.~Csisz\'ar and P. Narayan, ``The secret key capacity for multiple 
terminals,'' {\it IEEE Trans. Inform. Theory}, in review, 2003.

\end{itwreferences}

\end{document}